\documentclass[sigconf,screen,authorversion]{acmart}
\usepackage[utf8]{inputenc}        % ✓ Encoding of this TeX source file
\usepackage[nameinlink]{cleveref}  % ✓ Clever references

\copyrightyear{2024}
\acmYear{2024}
\setcopyright{acmlicensed}
\acmConference[ITiCSE 2024]{Proceedings of the 2024 Innovation and Technology in Computer Science Education V. 1}{July 8--10, 2024}{Milan, Italy}
\acmBooktitle{Proceedings of the 2024 Innovation and Technology in Computer Science Education V. 1 (ITiCSE 2024), July 8--10, 2024, Milan, Italy}
\acmDOI{10.1145/3649217.3653563}
\acmISBN{979-8-4007-0600-4/24/07}

\begin{document}
% Using \vspace to manually adjust vertical spacing is not allowed.

%% The "title" command has an optional parameter to define a "short title" for page headers.
\title{Comparison of Three Programming Error Measures for~Explaining Variability in CS1 Grades}
% Do not insert line breaks in your title.

\author{Valdemar Švábenský}
\orcid{0000-0001-8546-280X}
\email{valdemar.research@gmail.com}
\affiliation{
  \institution{University of Pennsylvania}
  \city{Philadelphia}
  \state{PA}
  \country{USA}
}

\author{Maciej Pankiewicz}
\orcid{0000-0002-6945-0523}
\email{maciej_pankiewicz@sggw.edu.pl}
\affiliation{
  \institution{Warsaw University of Life Sciences}
  \city{Warsaw}
  \country{Poland}
}

\author{Jiayi Zhang}
\orcid{0000-0002-7334-4256}
\email{joycez@upenn.edu}
\affiliation{
  \institution{University of Pennsylvania}
  \city{Philadelphia}
  \state{PA}
  \country{USA}
}

\author{Elizabeth B. Cloude}
\orcid{0000-0002-7599-6768}
\email{elizabeth.cloude@tuni.fi}
\affiliation{
  \institution{Tampere University}
  \city{Tampere}
  \country{Finland}
}

\author{Ryan S. Baker}
\orcid{0000-0002-3051-3232}
\email{ryanshaunbaker@gmail.com}
\affiliation{
  \institution{University of Pennsylvania}
  \city{Philadelphia}
  \state{PA}
  \country{USA}
}

\author{Eric Fouh}
\orcid{0000-0003-3869-9112}
\email{efouh@cis.upenn.edu}
\affiliation{
  \institution{University of Pennsylvania}
  \city{Philadelphia}
  \state{PA}
  \country{USA}
}

%% Define a concise list of authors' names for page headers.
%\renewcommand{\shortauthors}{Anonymous author(s)}

%% We have 249 out of 250 words
\begin{abstract}
Programming courses can be challenging for first year university students, especially for those without prior coding experience. Students initially struggle with code syntax, but as more advanced topics are introduced across a semester, the difficulty in learning to program shifts to learning computational thinking (e.g., debugging strategies). This study examined the relationships between students' rate of programming errors and their grades on two exams. Using an online integrated development environment, data were collected from 280 students in a Java programming course. The course had two parts. The first focused on introductory procedural programming and culminated with exam 1, while the second part covered more complex topics and object-oriented programming and ended with exam 2. To measure students' programming abilities, 51095 code snapshots were collected from students while they completed assignments that were autograded based on unit tests. Compiler and runtime errors were extracted from the snapshots, and three measures -- Error Count, Error Quotient and Repeated Error Density -- were explored to identify the best measure explaining variability in exam grades. Models utilizing Error Quotient outperformed the models using the other two measures, in terms of the explained variability in grades and Bayesian Information Criterion. Compiler errors were significant predictors of exam 1 grades but not exam 2 grades; only runtime errors significantly predicted exam 2 grades. The findings indicate that leveraging Error Quotient with multiple error types (compiler and runtime) may be a better measure of students' introductory programming abilities, though still not explaining most of the observed variability.
\end{abstract}

%% The code below is generated by http://dl.acm.org/ccs.cfm.


%% Pick words that accurately describe the work. Separate with commas for PDF and semi-colons for the ACM web form.

\maketitle

% ==================== Section start ====================
\section{Introduction}

A key component of computing education is acquiring knowledge, skills, and abilities (KSAs) related to programming. However, introductory programming courses are notoriously difficult to master for many undergraduates~\cite{Quille2019cs1}, and students' deficiencies in KSAs manifest in various struggles and errors they make while solving programming assignments.

In programming courses, students typically use an integrated development environment (IDE) to complete practical assignments, which are evaluated by an automated assessment tool. Data from the IDE and the automated assessment can be used by educators and researchers to determine students' KSAs and guide instruction~\cite{ihantola2015educational}.

\subsection{Outcome-based vs. Behavioral Measures}

To quantify students' KSAs, researchers often use static \textit{outcome-based measures} and summative assessments, such as scores and grades~\cite{zhang2022exploring, xie2019item, santos2020systematic}. However, these data capture only the final submitted form of the assignment, which a student may have iteratively refined to make it optimal using the results of the automated assessment. Therefore, relying only on outcome-based data to determine KSAs misses information on the process and students' approach to solving the programming assignment, such as differences in terms of time or debugging sessions prior to submitting the final product.

Another method for understanding KSAs in programming involves instrumenting the IDE to collect process-level data on students' learning behaviors during coding~\cite{azcona2019detecting, fields2016combining, carter2017using}. The resulting \textit{behavioral measures}, such as the number of recurring errors that students made while completing an assignment, may be better for measuring KSAs. These measures express how a student attempted to solve an assignment or debug an error based on changes made to their code. Specifically, error measures capture process-level information on the changes (or lack thereof) in the correctness of code. Behaviors related to programming errors also indicate students' ability to identify and debug an error effectively, which significantly impacts one's success in solving a programming problem \cite{ihantola2015educational}.

\subsection{Goals and Scope of This Paper}

Our literature review 
(see \Cref{sec:lit-review}) showed that it is unclear which programming error measure is the best at capturing student KSAs. Therefore, this study aims to build statistical models that use behavioral error measures to explain variability in outcome-based measures. We examine relationships between students' (a) rate of compiler and runtime errors across multiple programming assignments that vary in difficulty and (b) performance on two exams that capture programming knowledge relevant to the assignments.

We define and compare the rate of the errors using three different measures: Error Count, Jadud's Error Quotient, and Repeated Error Density. Our study is guided by the following research questions:
\begin{itemize}
    \item \textit{Which of the three error measures best explains variability in outcome-based measures (course exam grades)?}
    \item \textit{Do the results change when using only data about students' compiler errors vs. when adding data about runtime errors?}
\end{itemize}

% ==================== Section start ====================
\section{Review of Related Literature}
\label{sec:lit-review}

This section starts with an overview of studies that analyze programming errors to determine student performance (\Cref{subsec:related-work-performance}). Next, \Cref{subsec:related-work-measures} provides definitions and usage of specific error measures. \Cref{subsec:related-work-novelty} summarizes the novelty of this paper.

\subsection{Programming Errors and Performance}
\label{subsec:related-work-performance}

Relationships between students' programming errors and their performance are an area of substantial interest for computing educators and researchers (e.g., \cite{jadud2006exploration, tabanao2011predicting, Brown2018blackbox}). In a classic work, \citet{pea1986language} argued that programming errors result from misconceptions about topics held by programmers. \citet{ko2005framework} offered a more formal framework for understanding errors, arguing that errors do not always result from cognitive failures on the part of the programmer, but may stem from a variety of external factors, such as problems with the IDE or work interruptions.

Other researchers focused on relationships between performance and error debugging. \citet{denny2012all} conducted a study to explore the time it took students to debug common syntax errors. Their results showed that the higher-performing students do not debug common errors more quickly than low performers. Instead, certain types of common syntax errors required a significant amount of time for all students to debug. Similar findings were reported by \citet{rodrigo2013analysis} in a different teaching context. 

\subsection{Error Measures in Programming Education}
\label{subsec:related-work-measures}

This section defines two programming error measures\footnote{We use the term \textit{error measures} to be consistent with the cited literature. However, since they capture only errors reported in the IDE (which are a result of student error), a more accurate term might perhaps be \textit{error message measures}. Since some types of student errors are not recognized by these measures, they are effectively an (imperfect) proxy for student errors.} and reviews their applications in previous research.

\subsubsection{Jadud's Error Quotient (EQ)}
The EQ is one of the most studied measures of programming errors. It uses compilation log data to quantify the degree of repeated compiler errors in consecutive compilation events~\cite{jadud2006methods}. The EQ value is a decimal number between 0 and 1, indicating students' struggle to solve the programming problem. A value of 0 suggests that a student had only successful or a mix of successful and unsuccessful compilations. The value increases for each pair of successive compilation events that ended in a compiler error; additional penalty is applied if the error type was the same in both of those events. A maximum value of 1 suggests the student consistently encountered the same type of error.

\citet{jadud2006methods} indicated that a higher EQ is negatively associated with learning outcomes. The results revealed a significant, but weak negative relationship between a student's EQ and their average grade on programming assignments ($R^2=0.11$). Similarly, EQ explained only 25\% of the variability in final course grades, suggesting other factors may be playing a role in capturing this information.

To further explore relationships between EQ and grades, \citet{rodrigo2013analysis} conducted a mixed-methods study of differences in the effectiveness of debugging compilation errors between high-, average-, or low-performing students on a midterm exam. The EQ was used to define how effective students identified and debugged the errors. The qualitative analysis showed that high-performing students on the midterm perceived they had an easier time debugging compilation errors, compared to the average and low performers. However, the quantitative analysis revealed no differences in EQ scores between the groups. This result indicated that all the students struggled equally to debug the errors, regardless of their midterm grades and perceived abilities. Therefore, while some students perform better on exams, it may not reflect their actual ability to effectively debug the compilation errors in their programs.

These results might be explained by the fact that EQ accounts only for the frequency of errors that students make, but omits other aspects such as whether the error reoccurs over time or the error severity. Thus, some types of errors may have minimal impact on a student's performance in programming but could occur frequently, whereas less frequent errors could have a more detrimental impact on a students' knowledge and performance in programming.

In addition, empirical findings suggest that the EQ may vary by different groups, contexts, and environments. \citet{jadud2015aggregate} presented evidence of this property in a large-scale study of 27698 programmers. The EQ score was calculated for each student, and those who differed by country had significantly different EQ scores.

Finally, EQ cannot fully represent variability in different error sequences. Consider the following hypothetical scenario with two learners, A and B. Learner A encounters two successive compilations with the same error, yielding an EQ value of 1. Conversely, Learner B, who encounters the same error in three consecutive compilations, also has an EQ of 1. Furthermore, consider a third student, C, who undergoes two distinct series of two consecutive compilations with repeated errors. Yet again, the EQ remains at 1.

\subsubsection{Repeated Error Density (RED)}
The RED is a more generalizable measure formulated to capture some of the information omitted by the EQ~\cite{becker2016new}. The RED captures differences between students' errors at a more granular level. Specifically, it sums up a score calculated on each repeated error encountered in a sequence of consecutive compilation events. This way, RED accounts for not only the number of repeated errors, but also the number of series in which those repeated errors emerge and the lengths at which these errors continue to occur until fixed. As a result, \citet{becker2016new} argues that compared to EQ, RED is less influenced by the context. However, like EQ, RED does not consider the type of error that students made, which can vary in terms of severity.

RED's minimum value is 0 if the programming session was without a pair of errors in consecutive compilations. The maximum value of this measure is not constrained. A higher value denotes a higher level of struggle with the same error in successive compilations snapshots. Applying the RED measure to the scenario above, the values for students A, B, and C are 0.5, 3.2, and 1, respectively.

\citet{becker2016new} also compared EQ and RED using two data sets of compiler logs of novice Java programmers. Students were randomly assigned to one of two conditions: intervention, whose IDE provided enhanced error messages, whereas students in the control condition received normal error messages from the IDE. Students in the intervention group had significantly fewer errors overall at the group level, and similarly to the EQ, the RED scores were lower at the student level, providing evidence about validity of RED. 

\subsubsection{Summary}
While EQ is valuable for assessing programming errors, it cannot differentiate error sequences and might be sensitive to context. RED addresses some of these limitations by offering a more granular analysis of error sequences, but its properties were not studied as extensively as in the case of EQ.

More research is needed to evaluate the relationships between EQ/RED scores and student performance, specifically, assessing the degree to which EQ/RED can explain variability in performance outcomes. This is crucial for achieving a better understanding of the strengths and limitations of each measure in capturing relevant information about students' debugging processes and programming KSAs. This understanding, in turn, can serve to inform instructional strategies, curriculum development, and improve the effectiveness of programming education.

\subsection{Our Research Contributions}
\label{subsec:related-work-novelty}

The review of related work shows that different error measures work better in different contexts. The empirical results were mixed across the different studies, as to how each score related to performance. Therefore, this study brings the following contributions to computing education research:
\begin{itemize}
    \item We compared three error measures (EQ, RED, and a baseline Error Count measure) to determine which best predicts grades. Previous work mostly used only one measure; comparative studies of multiple measures are rare~\cite{becker2016new, tablatin2020relationship, qian2020investigation}.
    \item We extended prior work by including measures of both compiler and runtime errors. Previous work employed mostly measures of compiler errors. Runtime errors were not used much~\cite{carter2017using, Carter2015}, but may indicate deeper insights about students' problem-solving abilities since compiler errors typically stem from syntax errors that are relatively simple to fix.
    \item We replicated the findings of previous research on error measures in a different teaching context.
\end{itemize}

% ==================== Section start ====================
\section{Research Methods}

\begin{figure}[t]
\centering
\includegraphics[width=\linewidth]{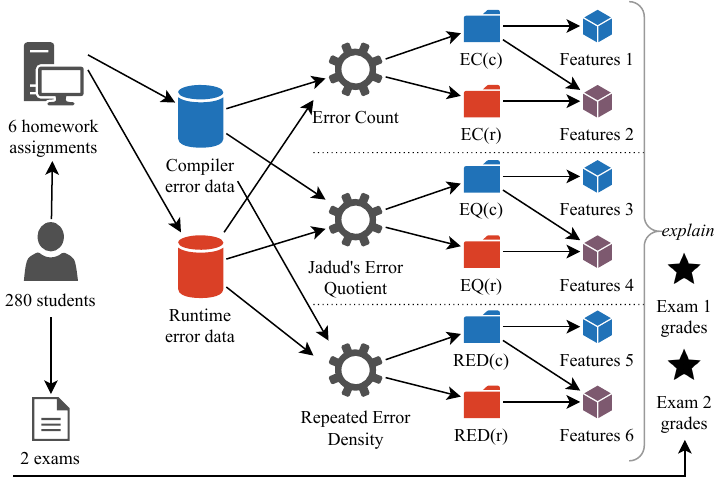}
\caption{High-level overview of the design of this study.}
\Description{Diagram illustrating the paragraph at the beginning of Section 3.}
\label{fig:design}
\end{figure}

\Cref{fig:design} represents the study design. Novice programming students ($n=280$) completed 8 homework assignments (6 of which were used in the study, see \Cref{subsec:methods-course}) and 2 exams in an introductory programming course. Data on compiler and runtime errors were collected from the IDE during solving the homework assignments, and 3 error measures were calculated to quantify the degree of student error. Next, we examined relationships between error measures and exam grades to determine which error measure was the best predictor of performance outcomes in the course. The following sections provide details on the individual aspects of the study.

\subsection{Course Design and Content}
\label{subsec:methods-course}

The students who participated in this study were enrolled in the CS1 course at University of Pennsylvania (a large, highly selective U.S. university) during Fall 2020. The course was taught in Java and spanned a standard 14-week semester, which ran partially online due to the COVID-19 restrictions at that time.

\subsubsection{Taught Topics}
The first part of the course covered procedural programming and introductory concepts, such as variables and data types, conditionals and loops, functions, arrays, and recursion. The second part of the course focused on object-oriented programming, unit testing, and abstract data types.

\subsubsection{Programming Assignments}
Throughout the semester, students had to complete eight programming homework (HW) assignments. All assignment grades were aggregated equally to account for a total of 60\% of the final course grade. The assignments were labeled from 1 to 8, and six of them (3--8) were considered in this study. HW 1 could not be anonymized since students were asked to include identifiable information in their program and it involved the graphical user interface; and HW 2 was not autograded; so these two were excluded from the analysis.

For each analyzed assignment, students had an unlimited number of submissions. With each submission, they received immediate feedback based on the results of unit tests, which reported the name of the test and a pass/fail outcome. The unit tests were created by the teaching staff, and students did not see the unit test definitions.

\subsubsection{Exams}
Students also had to complete 2 timed exams (22\% of the grade). The first exam was administered towards the middle of the term after HW 4. The second exam (non-cumulative final) took place near the end of the term after all HW assignments were completed. Exam~1 focused only on the introductory programming topics covered in HW 1--4. Exam 2 focused more acutely on the topics in HW 5--8 but also built on KSAs from HW 1--4. In this paper, we used data about student errors in the autograded HW 3--8 to explain their grades from exams 1 and 2.

\subsubsection{Other Parts of the Grading Scheme}
The remaining 18\% of the grade was based on attendance, code reviews, and online quizzes. These data were not used in this paper since our goal was to measure learning outcomes (expressed by grades) using predictor variables based on programming assignments.

\subsubsection{Tools Used in the Course}
Students completed programming assignments in Codio~\cite{codio}, an online IDE. Codio also housed the course's lecture notes, which served as a textbook. Programming assignments were submitted to Gradescope~\cite{gradescope}, an online automatic grading platform that also delivered the reading quizzes. 
 
\subsection{Student Population and Research Ethics}
During the Fall 2020 semester, 301 students were enrolled in the CS1 course, out of which 280 completed both exams. Only these students were included in our analysis.

Students in the course did not have prior computing experience and were predominantly in their first semester of undergraduate studies, with undeclared major. We neither collected nor had access to demographic data due to the university's data protection policy.

Before running the study, the institutional review board (IRB) determined the study to be exempt. All data were anonymized to maintain students' confidentiality and privacy.

\subsection{Data Collection and Preprocessing}

An internal script was used to enable the collection of data on students' programs in the Codio IDE. As students completed the homework assignments, the script collected a snapshot of each student’s code in Codio after every period of inactivity. Inactivity was defined as not interacting with the IDE for more than 10 minutes.

After raw snapshot data were collected, each snapshot was checked to determine whether it satisfied all the requirements of an assignment (e.g., all required files were present with correct file names). Although some invalid snapshots were occurring at the beginning of a student's work on an assignment, the vast majority of the snapshots satisfied the requirements. 

Each valid snapshot was then evaluated by an autograder. First, it was compiled to identify \textit{compiler errors}. If the snapshot compiled successfully, then the code was executed against a suite of unit tests to identify \textit{runtime errors}, which are Java exceptions thrown while running a test (not an outcome of a test). As a result, each snapshot ended up in one of three possible states:
\begin{itemize}
    \item Did not compile (had one or more compiler errors). If multiple compiler errors occurred, all these errors were reported to the student as well as counted for our analysis.
    \item Compiled successfully but had one or more runtime errors. For some HW assignments, testing stopped when the first runtime error was thrown, while for other assignments, the test suite executed the tests in parallel.
    \item Compiled successfully and finished without a runtime error.
\end{itemize}

The compiler and runtime error data from valid snapshots were included in our analysis. \Cref{tab:data} reports the totals of collected data types from all students. All data used in this research, along with more detailed descriptive statistics, are available (see \Cref{subsec:conclusion-materials}).

\begin{table}[h]
\caption{Counts of the individual types of the collected data. (HW 6 had much more snapshots since it was newly introduced that semester, so students made more attempts.)}
\label{tab:data}

\begin{tabular}{l|rrrrr}

HW \#      & Snap- & Compiler & Runtime & Failed & Passed \\[-1mm]
(students) & shots & errors   & errors  & tests  & tests  \\ \hline

\#3 (295) &  5763 &  1969 &  1511 &  28160 &   40123 \\
\#4 (281) &  2898 &  1074 &   863 &  24057 &   26919 \\
\#5 (281) &  6778 &  1314 & 18763 &  96933 &  132513 \\
\#6 (279) & 22450 &  2424 & 27525 &  55523 &  765543 \\
\#7 (280) &  5192 &  1351 &  8549 &  49897 &   71376 \\
\#8 (278) &  8014 &  2551 &  8192 &  21999 &  204741 \\ \hline
Total     & 51095 & 10683 & 65403 & 276569 & 1241215 \\

\end{tabular}
\end{table}

\subsection{Three Chosen Error Measures}

This section details the error measures, which are summarized in \Cref{tab:measures}. We chose these measures because they gradually increase in complexity, building on top of each other. Then, we describe computing the values of each measure for further analysis.

\begin{table}[h]
\caption{Summary of the three error measures. For all of them, a higher value indicates a higher rate of student error.}
\label{tab:measures}

\begin{tabular}{l|lll}

Measure & Value type             & Range                  & Used in \\ \hline

EC  & integer ($\mathbb{N}_0$)   & $0, 1, \ldots, \infty$ & --- \\
EQ  & decimal ($\mathbb{R}_0^+$) & $[0, 1]$               & \cite{jadud2006methods, rodrigo2013analysis, jadud2015aggregate, tablatin2020relationship, qian2020investigation} \\
RED & decimal ($\mathbb{R}_0^+$) & $[0, \infty)$          & \cite{becker2016new, tablatin2020relationship} \\

\end{tabular}
\end{table}

\subsubsection{Error Count (EC)}
Error Count is simply the total number of either compiler or runtime errors a student made in all snapshots. It is a non-negative integer, i.e., it ranges from 0 to potentially infinity. We use EC as the simplest error measure to obtain a baseline for further comparison with two other error measures: EQ and RED.

\subsubsection{Jadud's Error Quotient (EQ)}
The EQ was introduced in \Cref{subsec:related-work-measures}. To obtain EQ values for our compiler error data, we wrote a Python script that implemented the published algorithm for computing the EQ~\cite{jadud2006methods}. Although EQ was originally defined only for compiler errors, we extended this approach to also process runtime errors to obtain a comparison to values for compiler errors.

\subsubsection{Repeated Error Density (RED)}
The RED was also introduced in \Cref{subsec:related-work-measures}. To obtain the RED value for each student~\cite{becker2016new}, we created a data processing script using RapidMiner and Python. The process to obtain the RED values was analogous to EQ.

\subsection{Regression Statistical Analysis}

Our research goal was to understand and compare relationships between error measures (computed on compiler and runtime errors) and students’ performance outcomes (expressed by grades from exams 1--2), to identify a measure that best determines performance. To achieve this goal, a series of regression analyses were performed.

\subsubsection{Choice of the Regression Method}
Grades from the two exams were left-skewed, with the median grades being 84\% and 83\%. Because of this non-normal data distribution, a non-parametric rank-based regression~\cite{Kloke2012rfit} was used to explain exam grades.

\subsubsection{Choice of Feature Variables}
\Cref{fig:design} shows the six types of feature variables used in our models. The variables define values from the corresponding homework assignments assigned up to the date of the exam: HW 3--4 for exam 1, and HW 3--8 for exam 2.

To understand the differences between individual error measures and find the best one, we built separate models with predictors for each measure alone: either EC, EQ, or RED. As such, we did not build models that included a combination of predictors from different error measures since all the variables relied on a measure of error.

In addition, we compared the usage of compiler errors alone with the additional inclusion of runtime errors. We did not build models employing runtime errors alone, because runtime errors appear less frequently in general (if the code compilation fails on a compiler error, there is no possibility for a runtime error to occur).

To summarize, this resulted in 12 regression models altogether (2 exams $\times$ explained by 3 error measures $\times$ using 2 types of errors).

For example, the first model in \Cref{tab:results} predicted exam 1 grades using EC for compiler errors. This means that we collected the counts of compiler errors students made in HW 3 and HW 4, and used these two values to predict students' performance on exam 1. 

Lastly, as a ``benchmark'' comparison, homework grades were used as feature variables to predict students' grades on the two exams. We did not build models combining both homework grades and error measures as predictor variables, because we aimed to explore and compare the utility of the error measures alone.

\subsubsection{Implementation}
All steps of the regression analysis were implemented in R using the \texttt{Rfit} package (latest available version 0.24.2). The code used in this research is available (see \Cref{subsec:conclusion-materials}).

\subsubsection{Model Evaluation}
Each of the 12 models was evaluated based on the standard metrics listed in \Cref{tab:results}. We also used the Bayesian information criterion, BIC'~\cite[Equation 26]{Raftery1995bayesian}, which is adjusted with respect to the number of predictor variables. The more negative the BIC' value, the better. Moreover, if the difference between BIC' of two models is greater than 6, it is strong evidence that a model with the lower BIC' is significantly better~\cite[Table 6]{Raftery1995bayesian}.

% ==================== Section start ====================
\section{Results and Discussion}

\subsection{Explanation of Exam Grades}

\begin{table*}[t]
\caption{Regression analysis results. Measure: EC = Error Count, EQ = Jadud's Error Quotient, RED = Repeated Error Density. Predictors: $c_k$ or $r_k$ = value of the error measure for compiler or runtime errors for homework $k$; $hw_k$ = number of points for homework $k$. Statistical significance (p-value for the predictor variables and models): (***) < 0.001 < (**) < 0.01 < (*) < 0.05.}
\label{tab:results}

\begin{tabular}{|l||l|r|r|r||l|r|r|r|}
\hline

\hfill Value: & \multicolumn{4}{c||}{\textbf{Exam 1} (HW 3--4), $n=280$ students} & \multicolumn{4}{c|}{\textbf{Exam 2} (HW 3--8), $n=280$ students} \\ \cline{2-9}
Predicted by:                     & Sig. predictors        & Model F     & $R^2$ & BIC'     & Sig. predictors                        & Model F     & $R^2$ & BIC'     \\ \hline \hline
\textbf{EC (compiler)}            & $c_3$ (**), $c_4$ (*)  &  9.81 (***) & 0.086 & $-$6.02  & none                                   &  2.50 (**)  & 0.099 & 2.07     \\
\textbf{EC (compiler + runtime)}  & $c_3$ (**), $c_4$ (*)  &  5.63 (***) & 0.098 & $-$2.78  & $c_6$ (*), $r_5$ (***), $r_8$ (**)     &  3.35 (***) & 0.235 & $-$3.15  \\ \hline
\textbf{EQ (compiler)}            & $c_3$ (***)            & 30.38 (***) & 0.181 & $-$19.39 & $c_3$ (***), $c_6$ (*)                 & 10.40 (***) & 0.190 & $-$10.95 \\
\textbf{EQ (compiler + runtime)}  & $c_3$ (***)            & 15.46 (***) & 0.185 & $-$15.04 & $c_3$ (***), $c_6$ (*), $r_8$ (***)    &  7.73 (***) & 0.264 & $-$7.83  \\ \hline
\textbf{RED (compiler)}           & $c_3$ (***), $c_4$ (*) & 15.91 (***) & 0.105 & $-$8.65  & $c_4$ (*)                              &  3.44 (**)  & 0.072 & 5.59     \\
\textbf{RED (compiler + runtime)} & $c_3$ (***), $c_4$ (*) &  8.61 (***) & 0.114 & $-$4.91  & $r_5$ (***), $r_8$ (***)               &  5.09 (***) & 0.190 & 3.69     \\ \hline
\textbf{HW grades (benchmark)}    & $hw_3$ (***)           & 46.79 (***) & 0.257 & $-$31.23 & $hw_3$ (***), $hw_5$ (*), $hw_6$ (***) & 21.91 (***) & 0.330 & $-$34.02 \\

\hline
\end{tabular}

\end{table*}

\Cref{tab:results} reports the rank-based regression models with each of the three measures, predicting students’ grades for both exams. It also lists statistically significant predictors, as well as the models' $R^2$ and BIC' values. All significant predictors in the error-based models have a negative coefficient in the models, meaning that with lower error measure values, the exam grade improved, which is expected. 

\subsubsection{Overall Best Error Measure}

Based on both the values of $R^2$ and BIC', the EQ was the best for explaining variability in students' grades from both exams. This result held regardless of whether compiler or runtime error data were used. For exam 1, EQ was better than RED, which was better than EC. For exam 2, EQ was better than EC, which was better than RED. Still, the benchmark predictor (homework grades) outperformed all three error measures.

\subsubsection{Compiler vs. Runtime Errors}

When using compiler errors only, the best models reached $R^2=0.181$ for exam 1 and $R^2=0.190$ for exam 2. When adding the information about runtime errors too, both models improved: $R^2=0.185$ for exam 1 and $R^2=0.264$ for exam 2. However, this could be attributed to the fact that the model simply used more predictors. Regardless, both models were rather close to the benchmark prediction using homework grades.

\subsubsection{Exam 1 Grade}

Regarding exam 1 grades (covers HW 3--4), the significant predictors for all three error measures were $c_3$ and $c_4$, i.e., the variables based on compiler errors. Runtime errors were not a significant predictor in any model. Moreover, BIC' became worse when runtime errors were added for all three error measures.

We argue that this result is expected, since in the first part of the semester, the novice students struggled with syntax and were expected to make a lot of compiler errors. Therefore, it is not surprising that runtime errors do not make a big difference in the models on the first exam in the first part of the semester.

\subsubsection{Exam 2 Grade}

When predicting exam 2 grades (covers HW 3--8), variables based on runtime errors start appearing as significant predictors in all models. Unlike in exam 1, BIC' improved when runtime errors were added for two error measures (EC and RED). Compiler errors still remain relevant, but this is also because we again include data from HW 3--4 in the first part of the semester.

A possible explanation for this result is that as the semester progressed, most students became more familiar with Java syntax. Also, with the increased complexity of assignments and topics covered over the semester, runtime errors were more likely to occur and became more important in predicting students' performance. Arguably, runtime errors indicate higher-level misconceptions of students, which manifest later in the semester more strongly, unlike compiler errors that may occur due to trivial reasons such as typos.

\subsection{Comparison to the Related Research}

By splitting our target performance variable into two exams (one focused on basic topics, and the other on slightly more advanced topics within introductory programming) and collecting process measures on the errors made during programming, students' KSAs were captured at a granular level.

The explained variability in exam 1 grades was comparable to prior studies. EQ was the most suitable measure when used on compiler error data from introductory programming, despite some of the measure's limitations listed in \Cref{subsec:related-work-measures}. For EQ, our $R^2$ was between 18\% and 26\%, which was similar to or better than prior work: \citet{jadud2006methods} achieved 11\% on assignment grades and 25\% on final course grades, and \citet{tablatin2020relationship} achieved 20\% on midterm exam grades. In addition, \citet{tablatin2020relationship} achieved an $R^2$ of 12\% using RED, while our $R^2$ results for using RED ranged from 7\% to 19\%. Lower performance of RED is not surprising as it has been demonstrated to be effective particularly for short sequences of compilation events~\cite{becker2016new}.

The error measures did not work very well when compiler error data were used from intermediate-level topics (as in exam 2). Instead, applying the measures to runtime error data (contrary to their original use case) improved the models' explanatory power. 

\subsection{Limitations and Threats to Validity}

The results depend on the data collection method: snapshots were taken when a student remained idle, and the length of that pause was an arbitrary decision (we used a 10-minute threshold). An alternative would be to test various cut-offs or determine it based on the means and standard deviations of student pauses. Another common approach is to take snapshots per compilation attempt.

Plagiarism is a serious concern in programming courses and can compromise any study that uses assignment grades to answer research questions. To detect indicators of plagiarism, students' code was analyzed by the MOSS software~\cite{aiken2022moss}. We also used Codio's ``code playback''~\cite{codio-playback} feature to identify any large code paste indicating that the student received the code from a third party. Given that generative AI tools were not yet mature and open to the public at the time of the data collection, we can say that our rules around plagiarism remove it as a threat to validity of this study.

Finally, this study used data from a single course in one semester that ran partially online, so generalizations cannot be reliably made. However, since many other CS1 courses feature similar content and format, the findings could transfer beyond the original context.

\subsection{Open Research Challenges}

Utilizing information on compiler and runtime errors explained at most 26\% of the variability in students' exam performance. To improve this, there may be additional measures that should accompany error information. For example, information on strategies students use to debug different types of errors, their CS background, or their motivation could be a next step for future studies.

An additional direction for future work is to use fine-grained student data from HW assignments earlier in the semester to predict performance on exams later in the semester. These data may provide the diagnostic information to enable intervention and support.

Another research area is to identify topics and problems that students are struggling with in real-time to provide targeted help. For example, using item response theory with error measures may be a better way to determine topic difficulty based on students' rate of errors compared to outcome-based measures like grades.

% ==================== Section start ====================
\section{Conclusion}

Programming error measures are partially indicative of students' KSAs in introductory programming courses, but they do not explain most of the variability in exam grades. Our study evaluated EQ and RED, which were originally considered only for compiler errors, also in the context of runtime errors. The measures based solely on compiler errors were moderately useful as predictors of grades in the first part of the semester when students dealt with basic topics. However, in our course, where more complex topics were introduced in the second part of the semester, including runtime errors improved model performance later on. Overall, EQ always explained more variability in grades than RED or EC.

\subsection{Implications for Researchers and Educators}

Since the frequency of errors (compiler and runtime) affects student performance, it is important to help students understand them. However, error messages are unhelpful for most students~\cite{becker2019}. Therefore, the CS education community should continue to improve the pedagogical value of the feedback from error messages~\cite{Leinonen2023, phung2023}.

Developers of IDEs for introductory programming can enhance the IDEs with data collection and analysis of student errors. This may also support instructors' teaching practice in determining the KSAs of their students by predicting the exam outcomes and providing suitable interventions for students identified as struggling.

\subsection{Publicly Available Supplementary Materials}
\label{subsec:conclusion-materials}

The research data, software for data processing, and full results are available at \url{https://github.com/SERI-CS/iticse24-error-measures}.

\begin{acks}
This study was supported by the National Science Foundation (NSF; DUE-1946150). Any conclusions expressed in this material do not necessarily reflect the views of NSF.
\end{acks}

% Authors' names should be complete -- use full first names.
\bibliographystyle{ACM-Reference-Format}
\balance
\bibliography{references}

\end{document}